\documentclass[aps,pra,reprint,showpacs,superscriptaddress,10pt]{revtex4-1}
\usepackage{amsmath,amssymb,amsfonts,graphicx,bm,longtable,dcolumn,color,ulem}
\usepackage{mathrsfs}
\usepackage[utf8]{inputenc}
\usepackage{textcomp}
\usepackage[T1]{fontenc}
\usepackage{lmodern}

\begin{document}

\title{Radiative thermal rectification in many-body systems}

\author{Ivan Latella}
\email{ilatella@ub.edu}
\affiliation{Departament de F\'isica de la Mat\`eria Condensada, Universitat de Barcelona, Mart\'i i Franqu\`es 1, 08028 Barcelona, Spain}

\author{Philippe Ben-Abdallah}
\email{pba@institutoptique.fr}
\affiliation{Laboratoire Charles Fabry, UMR 8501, Institut d’Optique, CNRS, Université Paris-Saclay, 2 Avenue Augustin Fresnel, 91127 Palaiseau Cedex, France}

\author{Moladad Nikbakht}
\email{mnik@znu.ac.ir}
\affiliation{Department of Physics, University of Zanjan, Zanjan 45371-38791, Iran}

\begin{abstract}
Radiative thermal diodes based on two-element structures rectify heat flows thanks to a temperature dependence of material optical properties. The heat transport asymmetry through these systems, however, remains weak without a significant change in material properties with the temperature. Here we explore the heat transport in three-element radiative systems and demonstrate that a strong asymmetry in the thermal conductance can appear because of many-body interactions, without any dependence of optical properties on the temperature. The analysis of transport in three-body systems made with polar dielectrics and metallic layers reveals that rectification coefficients exceeding 50\% can be achieved in the near-field regime  with temperature differences of about 200\,K.  This work paves the way for compact devices to rectify near-field radiative heat fluxes over a broad temperature range and could have important applications in the domain of nanoscale thermal management.
\end{abstract}

\maketitle

\section{Introduction}

Near-field radiative heat transfer~\cite{PoldervH,LoomisPRB94,PendryJPhys1999,JoulainSurfSciRep05,VolokitinRMP07} between two bodies has attracted considerable interest in recent years. Theoretical developments~\cite{RodriguezPRL11,McCauleyPRB12,RodriguezPRB12,RodriguezPRB13,NarayanaswamyJQSRT14,MullerarXiv,BimontePRA09,KrugerPRB12,MessinaPRA11} and experimental measurements~\cite{KittelPRL05,NarayanaswamyPRB08,HuApplPhysLett08,ShenNanoLetters09,RousseauNaturePhoton09,OttensPRL11,KralikRevSciInstrum11,KralikPRL12,vanZwolPRL12a,vanZwolPRL12b,SongNatureNano15,KimNature15,StGelaisNatureNano16,KloppstecharXiv,WatjenAPL16} have confirmed the tremendous role played by near-field heat exchanges between solids at subwavelength separation distances and have highlighted the strong potential of this transfer in a wide range of applications such as energy harvesting~\cite{DiMatteoAPL2001,NarayanaswamyAPL2003,FiorinoNatNano2018,FiorinoNatNano2018,BhattNatComm2020}, heat assisted data recording~\cite{ChallenerNatPhot2009,StipeNatPhot2010}, near-field infrared spectroscopy~\cite{DeWildeNat2006,JonesNanoLett2012}, and active control of the heat exchange~\cite{BiehsAPL2011,HeOL2020,HeIJHMT2020}.
More recently, the near-field radiative heat transfer in a set of objects in mutual interaction was also investigated, for instance, considering a distribution of nanoparticles~\cite{BenAbdallahPRL11,MessinaPRB13,BenAbdallahPRL13,LanglaisOptExpress14,NikbakhtJAP14,BenAbdallahAPL06,BenAbdallahPRB08,NikbakhtEPL15,Tervo,NikbakhtPRB17,RamezanJAP2016} and structures with planar geometry~\cite{ZhengNanoscale11,MessinaPRL12,MessinaPRA14,LatellaPRAppl15,OrdonnezPRB2016,MessinaPRB2016,Latella_2017,LatellaPRB18,He19,Kan19,LatellaSciRep2020,ZolghadrPRB2020}.
These works have revealed the existence of specific effects that are attributed to many-body interactions and which offer new degrees of freedom to passively and actively manage heat currents at the nanoscale, in a way similar to the control of electric currents in solids. To date, many thermal analogues of electronics building blocks have been proposed to switch~\cite{LatellaPRL2017,PapadakisPRApplied2021}, focus~\cite{Ben-AbdallahPRL2019}, amplify~\cite{BenAbdallahPRL14,Ordonez2,LatellaPRAppied19}, and even split~\cite{Ben-AbdallahAPL2015,BenAbdallahPRL16,GuoACSPhot2020} heat fluxes using many-body systems. These systems have also been suggested to store thermal information~\cite{PBA-PRL2014-1,Dyakov,KhandekarAPL2017} and to make basic logical operations with heat~\cite{Ben-AbdallahPRB2016,KathmannSciRep2020}.

One of the fundamental building blocks in electronics is the diode, a two-terminal component with a highly asymmetric electric conductance which essentially conducts current in one direction. Thermal analogs of these devices have been proposed to rectify heat flux between two solids separated by a vacuum gap~\cite{OteyPRL2010}. This thermal rectification exploits the thermal dependence of optical properties of the materials in interaction~\cite{BasuAPL2011,IizukaJAP2012,WangNMTE2013,NefzaouiAO2014}, which naturally breaks the symmetry in the heat transport when the sign of the temperature gradient between the two solids is changed. In addition, a large asymmetry in the heat transport was reported in systems made with materials undergoing a phase change, such as metal-insulator transition materials~\cite{vanZwolPRB2011,Ben-AbdallahAPL2013,HuangIJHMT2013,YangAPL2013,GuIJMHT2015,YangJQSRT2015,GhanekarAPL2016,ZhengIJHMT2017,GhanekarOE2018,ChenAEM2021,ItoAPL2014,ItoNL2017,FiorinoACS2018,ForeroPRAppl2020} and normal metal-superconductor transition materials~\cite{NefzaouiAPL2014,OrdonezJAP2017,MoncadaPRAppl2021}. 
However, this strong rectification only occurs close to the critical temperature of these materials. In the present work, we show that many-body systems can be used to rectify heat fluxes over a broad temperature range. To evidence this fact, here we consider a simple system made up of three bodies separated by vacuum in which rectification can be achieved even when the temperature dependence of material properties is neglected. We demonstrate that the asymmetry of the thermal conductance is a direct consequence of many-body interactions in the system, and we illustrate the efficiency of such rectifiers with concrete examples.

The paper is organized as follows. After a phenomenological description of the rectification mechanism in Sec.~\ref{sec:rectifier}, we describe this effect in Sec.~\ref{sec:fluxes} using the rigorous framework of fluctuational electrodynamics. Results in concrete configuration are analyzed in Sec.~\ref{sec:rectification} for systems where the asymmetry is achieved by combining a polar material and a metal. Finally, a summary and concluding remarks are presented in Sec.~\ref{sec:conclusions}.

\section{Many-body thermal rectification: phenomenological approach\label{sec:rectifier}}

\begin{figure}
\includegraphics[scale=0.7]{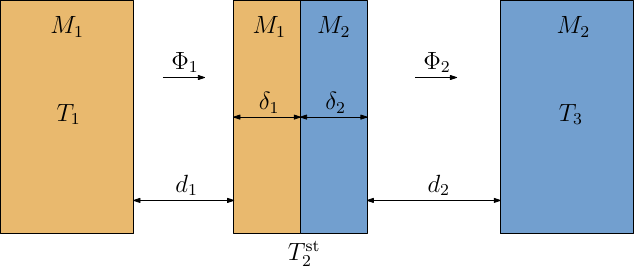}
\caption{Schematic representation of a three-body passive thermal rectifier. The device terminals at temperatures $T_1$ and $T_3$ (bodies 1 and 3, left and right, respectively) interact through an intermediate passive element (body 2, middle) at a stationary temperature $T_2^\mathrm{st}$. Two different materials, denoted as $M_1$ and $M_2$, are considered in order to introduce an asymmetry in the system. The quantities $\Phi_1$ and $\Phi_2$ represent the radiative energy flux in the two vacuum cavities with thicknesses $d_1$ and $d_2$.}
\label{sketch}
\end{figure}

The proposed rectification mechanism takes advantage of asymmetric radiative interactions between a passive intermediate body and two external terminals defining the direction of the heat flux.
As shown in Fig.~\ref{sketch}, the device consists of three bodies in which bodies 1 and 3 are assumed semi-infinite slabs at fixed temperatures $T_1$ and $T_3$, respectively, constituting the external terminals. Bodies 1 and 3 are made of different materials denoted as $M_1$ and $M_2$, respectively. 
To maximize the radiative coupling between the intermediate body and the left and right reservoirs, for this body (denoted as body 2) we consider a bilayer structure $M_1$-$M_2$ at temperature $T_2$ with thickness $\delta=\delta_1+\delta_2$, where $\delta_1$ and $\delta_2$ are the thicknesses of the constituting layers. Body 2 is separated by a distance $d_1$ from body 1 and by a distance $d_2$ from body $3$. Moreover, the temperature of body 2 here is taken as the stationary temperature $T_2=T_2^\mathrm{st}$ at which the net flux on this body vanishes, so the intermediate body passively interacts with the two external terminals.

The quantities $\Phi_1(T_1,T_2,T_3)$ and $\Phi_2(T_1,T_2,T_3)$ in Fig.~\ref{sketch} represent the energy flux in the two vacuum gaps, which are computed as the averaged Poynting vector in the direction perpendicular to the surfaces of the bodies (see Sec.~\ref{sec:fluxes}).
The stationary temperature $T_2=T_2^\mathrm{st}$ is then obtained by requiring that the heat flux is the same in the two cavities,
\begin{equation}
\Phi(T_1,T_2^\mathrm{st},T_3)\equiv\Phi_1(T_1,T_2^\mathrm{st},T_3) = \Phi_2(T_1,T_2^\mathrm{st},T_3).
\label{eq_temp}
\end{equation}
Under these conditions and taking $T_1=T_h$ and $T_3=T_c$ with $T_h>T_c$, the forward flux $J_f$ can be defined as the energy flux received by the terminal at temperature $T_3=T_c$,
\begin{equation}
J_f= \Phi(T_h,T_{f},T_c),
\label{forward_flux}
\end{equation}
where $T_{f}\equiv T_2^\mathrm{st}$ is the stationary temperature of body 2 in the forward scenario. The backward scenario corresponds to the situation in which the temperatures of the terminals are exchanged, taking $T_1=T_c$ and $T_3=T_h$. The backward flux $J_b$ can be obtained by exchanging the values of the temperatures $T_1$ and $T_3$ in the forward flux together with a global change in sign, so we can write
\begin{equation}
J_b=-\Phi(T_c,T_b,T_h),
\label{backward_flux}
\end{equation}
where $T_{b}\equiv T_2^\mathrm{st}$ is the stationary temperature of body 2 in the backward situation.
Furthermore, the rectification coefficient accounting for the system dependence on the direction of the heat flux is defined by
\begin{equation}
\eta= \frac{|J_f-J_b|}{\max(|J_f|,|J_b|)}.
\label{eta}
\end{equation}
If the system is fully symmetric, with materials $M_1=M_2$ and separations $d_1=d_2$, the rectification coefficient vanishes when the temperature dependence of the material properties is neglected. In contrast, we show below that an asymmetry in combination with many-body interactions can realize a nonvanishing rectification coefficient without any temperature dependence of the material properties.

It is worth noting that the stationary temperatures $T_f$ and $T_b$ in the forward and backward scenarios, respectively, do not coincide, in general, when the thermal symmetry is broken. We also point out that a trivial optical asymmetry is not enough to induce rectification when the temperature dependence of material properties is neglected. To highlight this fact with examples, let us consider a three-body system with heat fluxes following a power law of the form 
\begin{equation}
\Phi_1=\xi_1(T_1^\alpha-T_2^\alpha),\qquad \Phi_2=\xi_2(T_2^\alpha-T_3^\alpha).
\end{equation}
If, for instance, $\alpha=4$, we have blackbody radiation with $\xi_i=\sigma\epsilon_i$, where $\epsilon_1$ and $\epsilon_2$ are the emissivities which characterize the heat transfer asymmetry on both sides of the intermediate body, $\sigma$ being the Stefan-Boltzmann constant. If $\alpha=2$, we have a theoretical limit for the heat transfer in near-field regime between two dielectrics~\cite{PendryJPhys1999,VolokitinRMP07} with $\xi_i=k_B^2\kappa_i^2/24\hbar$, where $\kappa_1$ and $\kappa_2$ are some cut-off wave vectors characterizing the asymmetry in this case, $k_B$ and $\hbar$ being the Boltzmann constant and the reduced Planck constant, respectively. The stationary temperature $T_2=T_2^\mathrm{st}$ is defined by the condition $\Phi_1=\Phi_2$, so that
\begin{equation}
(T_2^\mathrm{st})^\alpha =\frac{\xi_1T_1^\alpha +\xi_2T_3^\alpha}{\xi_1 + \xi_2}.
\label{eq_temp_case}
\end{equation}
Under these conditions, the fluxes in the system can be written explicitly as a function of $T_1$ and $T_3$ only,
\begin{equation}
\Phi(T_1,T_3)=   \frac{\xi_1 \xi_2 }{\xi_1 + \xi_2}\left(T_1^\alpha -T_3^\alpha\right).
\end{equation}
Hence, in this case the forward flux $J_f=\Phi(T_h,T_c)$ coincides with the backward flux $J_b=-\Phi(T_c,T_h)$, leading to a vanishing rectification coefficient.
Notice that the stationary temperature (\ref{eq_temp_case}) is not invariant under the exchange $T_1\leftrightarrow T_3$, so this temperature is in general different in the forward and backward cases, even though there is no rectification.
Notice also that the same result applies if the $\xi_i$'s somehow depend on the separation distances, for instance through the cut-off wave vectors~\cite{Ben-AbdallahPRB2010}. These arguments demonstrate that a trivial optical asymmetry is not sufficient to induce rectification. To observe rectification, the asymmetry must introduce some effective temperature dependence on the transport properties of the system.

\section{Rigorous description of the rectification mechanism\label{sec:fluxes}}

The radiative heat exchange between macroscopic bodies can be rigorously described with the Landauer-like formalism.
In this case, the energy flux $\Phi_\gamma$ in the cavity $\gamma=1,2$ as sketched in Fig.~\ref{sketch} can be written as~\cite{MessinaPRL12,MessinaPRA14}
\begin{equation}
\Phi_{\gamma}=\int_0^\infty\frac{ d \omega}{2\pi} \hbar\omega \sum_{p}\int_0^\infty\frac{d  k }{2\pi}\,  k 
\left(  n_{ 1 2 }\hat{\mathcal{T}}^{ 1 }_{\gamma} + n_{ 2 3 }\hat{\mathcal{T}}^{ 2 }_{\gamma} \right),
\label{energy_flux}
\end{equation}
where $\omega$, $k$, and $p=\mathrm{TM,TE}$ are the frequency, wave vector parallel to the surfaces, and polarization of the electromagnetic radiation, respectively, while $n_{ij}=n_i-n_j$ set the difference of thermal distribution functions $n_j=1/(e^{\hbar\omega/k_BT_j}-1)$ at temperature $T_j$ associated with body $j$. 
In expression (\ref{energy_flux}), $\hat{\mathcal{T}}^j_\gamma$ are the Landauer-like coefficients, which define the energy transmission associated with mode $(\omega,k,p)$ through the system. For planar geometry, these coefficients are given by~\cite{Latella_2017} 
\begin{widetext}
\begin{equation}\begin{split}
\hat{\mathcal{T}}^{ 1 }_{2}
&= 
\frac{\Pi^{\text{pw}} |\tau_2 |^2 \big(1 -|\rho_1|^2\big)
\big( 1 -|\rho_3|^2 \big)}
{\big|1- \rho_1\rho_2^- e^{i2k_zd_1} \big|^2
\big|1- \rho_{12} \rho_3 e^{i2k_zd_2} \big|^2 } 
+ 
\frac{\Pi^{\text{ew}} 4 |\tau_2 |^2\text{Im}(\rho_1)
\text{Im}(\rho_3) 
e^{-2\mathrm{Im}(k_z)(d_1 +d_2)}}
{\big|1- \rho_1\rho_2^- e^{i2k_zd_1}\big|^2
\big|1- \rho_{12} \rho_3 e^{i2k_zd_2} \big|^2 },\\
\hat{\mathcal{T}}^{1}_{1}
&=    
\frac{\Pi^{\text{pw}} \big(1 -|\rho_1|^2\big)
\big( 1 -|\rho_{23}|^2 \big)}
{\big|1- \rho_1 \rho_{23} e^{i2k_zd_1}\big|^2}
+ 
\frac{\Pi^{\text{ew}} 4\text{Im}(\rho_1)
\text{Im}(\rho_{23})
e^{-2\mathrm{Im}(k_z)d_1} }
{\big|1- \rho_1 \rho_{23} e^{i2k_zd_1}\big|^2},\\
\hat{\mathcal{T}}^{2}_{2}
&=    
\frac{\Pi^{\text{pw}} \big(1 - |\rho_{12} |^2\big)
\big( 1 -|\rho_3|^2 \big)}
{\big|1- \rho_{12} \rho_3 e^{i2k_zd_2}\big|^2}
+ 
\frac{\Pi^{\text{ew}} 4\text{Im}(\rho_{12})
\text{Im}(\rho_3)
e^{-2\mathrm{Im}(k_z)d_2} }
{\big|1- \rho_{12} \rho_3 e^{i2k_zd_2}\big|^2},
\end{split}\label{coeff_seq}\end{equation} 
\end{widetext}
where $k_{z}=\sqrt{(\omega/c)^2- k^2}$ is the normal component of the wave vector in vacuum, and $\Pi^{\text{pw}}=\theta(\omega-ck)$ and $\Pi^{\text{ew}}=\theta(ck-\omega)$ are the propagating and evanescent wave projectors, respectively, with $c$ being the speed of light and $\theta(x)$ being the Heaviside step function. The remaining coefficient is obtained as $\hat{\mathcal{T}}^2_1 = \hat{\mathcal{T}}^1_2$, an equality which is valid for reciprocal media.
In Eqs.~(\ref{coeff_seq}) we have introduced the optical reflection coefficients $\rho_1$, $\rho_2^{+/-}$, and $\rho_3$ of bodies 1, 2, and 3, respectively, the optical transmission coefficient $\tau_2$ of body 2, and the two-body reflection coefficients given by
\begin{equation}
\rho_{12}= \rho_2^+ + \frac{(\tau_2)^2\rho_1e^{i2k_zd_1}}{1-\rho_1\rho_2^- e^{i2k_zd_1}}
\end{equation}
and
\begin{equation}
\rho_{23}= \rho_2^- + \frac{(\tau_2)^2\rho_3e^{i2k_zd_2}}{1-\rho_2^+\rho_3 e^{i2k_zd_2}} , 
\end{equation}
where the superscript $+$ indicates reflection to the right by body 2 and the superscript $-$ indicates reflection to the left (all these coefficients depend on $\omega$, $k$, and $p$). The dependence on the reflection direction in the coefficients $\rho_2^{+/-}$ accounts for the fact that the intermediate body is a bilayer, while transmission across this body does not depend on the direction of the transmitted field. Single-body reflection and transmission coefficients are given in the Appendix. As can be interpreted from Eq.~(\ref{energy_flux}), the coefficients $\hat{\mathcal{T}}_\gamma^\gamma$ quantify the transmission amplitude of modes participating in the heat exchange in cavity $\gamma$, while $\hat{\mathcal{T}}_\gamma^j$ with $j\neq\gamma$ quantifies the transmission of contributing modes from cavity $j$ to cavity $\gamma$.

Using Eqs.~(\ref{eq_temp}) and (\ref{forward_flux}) to express the forward flux such that $J_f= \frac{1}{2}[\Phi_1(T_h,T_{f},T_c)+\Phi_2(T_h,T_{f},T_c)]$ and taking into account Eq.~(\ref{energy_flux}), this flux can be written as
\begin{equation}
\begin{split}
J_f&=\frac{1}{2}\int_0^\infty\frac{ d \omega}{2\pi} \hbar\omega \sum_{p}\int_0^\infty\frac{d  k }{2\pi}\,  k \\
&\times \left(  n_{ h }\hat{\mathcal{T}}^{ 1 }_{1} - n_{ c }\hat{\mathcal{T}}^{ 2 }_{2}
-  n_{ f }[ \hat{\mathcal{T}}^{ 1 }_{1} - \hat{\mathcal{T}}^{ 2 }_{2}]  +n_{ hc }\hat{\mathcal{T}}^{ 1 }_{2}  \right),
\end{split}
\label{forward_flux_2}
\end{equation}
where $n_h$ and $n_c$ are the distribution functions evaluated at temperatures $T_h$ and $T_c$, respectively, for which $n_{hc}=n_h-n_c$, and $n_f$ is the distribution function evaluated at the stationary temperature $T_f$ of the intermediate body in the forward scenario. 
An expression analogous to Eq.~(\ref{forward_flux_2}) can be obtained for the backward flux. Considering materials with temperature-independent optical properties, the transmission coefficients do not depend on the scenario and we have
\begin{equation}
\begin{split}
J_f-J_b&=\frac{1}{2}\int_0^\infty\frac{ d \omega}{2\pi} \hbar\omega 
(n_{ h } + n_{ c } -n_f -n_b) \\
&\times\sum_{p}\int_0^\infty\frac{d  k }{2\pi}  k 
( \hat{\mathcal{T}}^{ 1 }_{1} - \hat{\mathcal{T}}^{ 2 }_{2} ) , 
\end{split}
\label{Jf_Jb}
\end{equation}
where $n_b$ is the distribution function evaluated at the stationary temperature $T_b$ of the intermediate body in the backward case. From Eq.~(\ref{Jf_Jb}) we note that an asymmetry in the efficiency of modes coupling between the subblocks of the system (i.e. $\hat{\mathcal{T}}^{ 1 }_{1}\neq \hat{\mathcal{T}}^{ 2 }_{2}$) can induce a rectification of the heat flux. This forces the system to be dissymmetric as well.  We also observe that the value of the equilibrium temperature of the intermediate body in the forward and backward scenarios plays a fundamental role in the dissymmetry of heat transport. 
In the next section we illustrate the many-body rectification by considering a simple system made with two materials of radically different nature, a polar material and a metal.

\section{Many-body rectification: some concrete examples\label{sec:rectification}}

\begin{figure}
\centering
\includegraphics[scale=0.9]{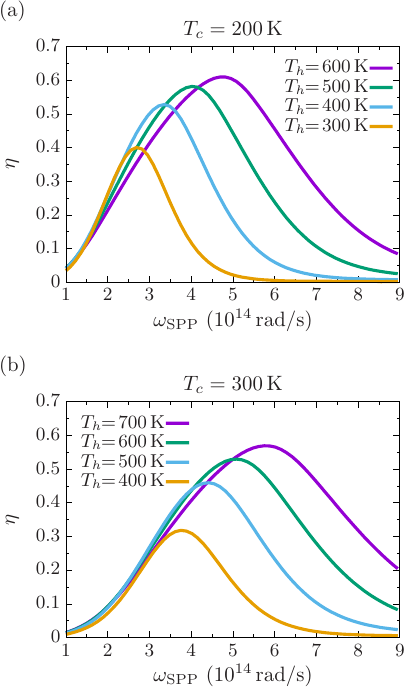}
\caption{Rectification coefficient for a system with materials $M_1=\mathrm{hBN}_\Omega$ and $M_2=\mathrm{Cu}$ as a function of the resonance frequency $\omega_\mathrm{SPP}$ supported by $\mathrm{hBN}_\Omega$. The cold temperature is set to $T_c=200\,$K in (a) and $T_c=300\,$K in (b). For low $\omega_\mathrm{SPP}$, the rectification increases for increasing $\omega_\mathrm{SPP}$ because this resonance is efficiently excited in the forward scenario. For high $\omega_\mathrm{SPP}$, the rectification decreases for increasing $\omega_\mathrm{SPP}$ because the overlap between the resonance and the distribution function evaluated at $T_h$ in the forward scenario is reduced.}
\label{fig_shift_Cu}
\end{figure}

As discussed in the previous section, the ability of the system to rectify heat flows is characterized by the asymmetry in the transmission coefficients $\hat{\mathcal{T}}^{ 1 }_{1}$ and  $\hat{\mathcal{T}}^{ 2 }_{2}$. Such an asymmetry can be realized by choosing the materials $M_1$ and $M_2$ constituting the cavities with resonances at different frequencies. Here we will consider these resonant frequencies to be well separated in the spectrum and take $M_1$ to be a polar material, whose resonance lies in the infrared and can be excited at the considered temperatures, and $M_2$ to be a metal, whose resonance is beyond the infrared and is not thermally excited in the working temperature range.

To describe the polar material, let us consider the Drude-Lorentz model for the permittivity
\begin{equation}
\varepsilon(\omega)=\varepsilon_\infty\frac{\omega_L^2-\omega^2-i\Gamma \omega}{\omega_T^2-\omega^2-i\Gamma \omega}, 
\label{Drude-Lorentz}
\end{equation}
where $\varepsilon_\infty$ is the high-frequency dielectric constant, $\omega_L$ is the longitudinal phonon frequency, $\omega_T$ is the transverse phonon frequency, and $\Gamma$ is the damping constant.
In order to study the influence of resonant waves in the rectification mechanism, we tune the optical properties of the material by introducing a shift $\Omega$ in the phonon frequencies, in such a way that the permittivity is written as
\begin{equation}
\varepsilon(\omega)=\varepsilon_\infty\frac{(\omega_L+\Omega)^2-\omega^2-i\Gamma \omega}{(\omega_T+\Omega)^2-\omega^2-i\Gamma \omega}. 
\label{eps_Omega}
\end{equation}
Taking into account this permittivity, the frequency $\omega_\mathrm{SPP}$ of the surface phonon polariton (SPP) supported by the material is given by $\mathrm{Re}[\varepsilon(\omega_\mathrm{SPP})]=-1$ and can be approximated as
\begin{equation}
\omega_\mathrm{SPP}\approx  \left[\frac{\varepsilon_\infty(\omega_L+\Omega)^2+(\omega_T+\Omega)^2}{1+\varepsilon_\infty}\right]^{1/2}.
\label{omega_SPP}
\end{equation}
When this resonance is thermally excited, the heat transfer in the near field is strongly dominated by the contribution of modes around this frequency. 
To define the material, here we take $\varepsilon_\infty=4.9$, $\omega_L=3.03\times10^{14}\,$rad/s, $\omega_T=2.57\times10^{14}\,$rad/s, and $\Gamma=1.0\times10^{12}\,$s$^{-1}$, which correspond~\cite{Eremets95} to hexagonal boron nitride (hBN) when $\Omega=0$. Hence, the material specified by the permittivity (\ref{eps_Omega}) will be denoted as hBN$_\Omega$. Furthermore, we consider copper (Cu) as the metal and describe its permittivity with the Drude model
\begin{equation}
\varepsilon(\omega)=\varepsilon_\infty -\frac{\omega_p^2}{\omega(\omega+i\Gamma)},
\end{equation}
where $\omega_p=1.12\times10^{16}\,$rad/s is the plasma frequency and we take $\varepsilon_\infty=1$ and $\Gamma=1.38\times10^{13}\,$s$^{-1}$~\cite{OrdalAO1895}.

\begin{figure*}
\centering
\includegraphics[scale=0.87]{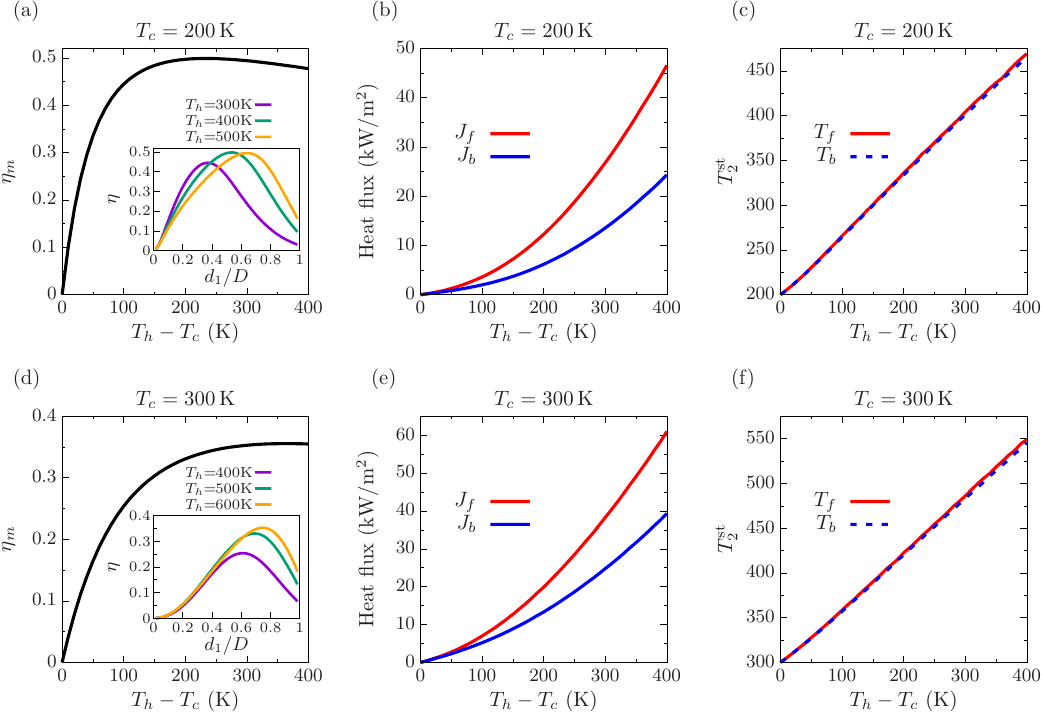}
\caption{Rectification for a system with $M_1=\mathrm{hBN}$ and $M_2=\mathrm{Cu}$. (a) Optimized rectification coefficient $\eta_m$ obtained by choosing the separation $d_1$ leading to maximum rectification for the considered temperatures, with $D=d_1+d_2=100\,$nm and cold temperature $T_c=200\,$K. The inset shows the dependence of the rectification coefficient $\eta$ on the separation distance for several hot temperatures $T_h$. The corresponding forward and backward fluxes $J_f$ and $J_b$, respectively, are shown in (b), while the stationary temperatures of the intermediate body $T_f$ and $T_b$ are plotted in (c). The same quantities are represented in (d)-(f), but taking the cold temperature as $T_c=300\,$K.}
\label{fig_hBN_Cu}
\end{figure*}

Following the scheme shown in Fig.~\ref{sketch}, we first consider materials $M_1=\mathrm{hBN}_\Omega$ and $M_2=\mathrm{Cu}$ with fixed separation distances $d_1=d_2=50\,$nm. For all configurations considered throughout the paper, the thicknesses of the layers in the intermediate body are set to $\delta_1=\delta_2=1\,\mu$m. We compute the rectification coefficient according to Eq.~(\ref{eta}) and plot it in Figs.~\ref{fig_shift_Cu}(a) and \ref{fig_shift_Cu}(b) as a function of $\omega_\mathrm{SPP}$ for $T_c=200\,$K and $T_c=300\,$K, respectively. The frequency $\omega_\mathrm{SPP}$ and the rectification coefficient are parametrized with the shift frequency $\Omega$, which in the figures varies from $-2\times10^{14}$ to $4\times10^{14}\,$rad/s.
Rectification factors of about $61\%$ and $57\%$ are observed in Figs.~\ref{fig_shift_Cu}(a) and \ref{fig_shift_Cu}(b), respectively, for a temperature difference of $T_h-T_c=400\,$K. Although the intrinsic temperature dependence of the optical properties has been neglected, rectification is achieved through the effective temperature dependence of the heat fluxes induced by many-body interactions. The main contribution to this temperature dependence arises because of the change in the overlap between the SPP supported by the polar material and the distribution function of this material when evaluated at $T_h$ and $T_c$. The rectification coefficient increases for increasing $\omega_\mathrm{SPP}$ because this resonance is efficiently excited in the forward scenario when the polar material is thermalized at $T_h$. As $\omega_\mathrm{SPP}$ increases beyond the Planck window defined by $T_h$, the rectification decreases because the SPP is not efficiently excited. Notice that the height and width of the rectification peaks in Figs.~\ref{fig_shift_Cu}(a) and \ref{fig_shift_Cu}(b) depend not only on the temperatures, but also on the properties defining the asymmetry such as permittivities of the two materials and separation distances.

For given materials and working temperatures, the performance of the rectifier can be optimized by taking the appropriate separation distances $d_1$ and $d_2$ between the bodies. We now consider $D=d_1+d_2$ fixed and take $D=100\,$nm. Under these conditions, the rectification coefficient can be seen as a function $\eta(d_1)$ and we are interested in the optimum value $\eta_m$ such that $\eta_m=\max_{d_1}\eta(d_1)$. Here we consider materials $M_1=\mathrm{hBN}$, described by the permittivity (\ref{Drude-Lorentz}), and $M_2=\mathrm{Cu}$. Taking $T_c=200\,$K, in Fig.~\ref{fig_hBN_Cu}(a) we plot the optimized rectification coefficient $\eta_m$ for this configuration as a function of the temperature difference $T_h-T_c$. A rectification of about $50\%$ is observed for a temperature difference close to $200\,$K. In the inset of this figure, we show $\eta(d_1)$ for several hot temperatures $T_h$, highlighting the dependence of the rectification coefficient on the separation distances. 
As can be seen in this inset, on the one hand, when $d_1/D$ approaches zero, meaning that the hBN surfaces of body 1 and body 2 are close to contact, a strong thermalization~\cite{LatellaSciRep2020} mediated by the SPP resonances enforces that a two-body system is recovered in this limit. Since the temperature dependence of material properties is neglected here, the rectification coefficient vanishes in this case. On the other hand, when the Cu surfaces of body 2 and body 3 are close to each other with $d_1/D$ approaching unity, the temperature difference $T_2^\mathrm{st}-T_3$ between these bodies is small but finite~\cite{LatellaSciRep2020}, since only radiative channels for heat exchange are considered and no resonances are excited in metals at the working temperatures. Thus, a nonzero rectification coefficient can be observed at small separations, provided other channels for heat exchange are not active. At contact, thermalization is driven by conduction and the two-body behavior with a vanishing rectification coefficient is recovered.

The forward and backward fluxes corresponding to the optimal coefficient $\eta_m$ in Fig.~\ref{fig_hBN_Cu}(a) are plotted in Fig.~\ref{fig_hBN_Cu}(b), while the stationary temperatures of the intermediate body $T_f$ and $T_b$ in the forward and backward scenarios, respectively, are shown in Fig.~\ref{fig_hBN_Cu}(c). Interestingly, maximum rectification is achieved for separation distances yielding similar values of $T_f$ and $T_b$. This behavior contrasts with that obtained with trivial asymmetries as discussed in Sec.~\ref{sec:rectifier}, in which asymmetric configurations always lead to different $T_f$ and $T_b$, but the rectification coefficient identically vanishes. The optimized rectification coefficient $\eta_m$ for the same configuration but assuming $T_c=300\,$K is shown in Fig.~\ref{fig_hBN_Cu}(d). In this case, the rectification coefficient can reach values up to $35\%$ for temperature differences $T_h-T_c$ above 200\,K. As compared with the previous case in which $T_c=200\,$K, the reduction in the rectification coefficient can be attributed to the fact that the SPP supported by hBN is more efficiently excited in the backward situation with $T_c=300\,$K, so the difference of fluxes in the forward and backward scenarios is relatively smaller. The fluxes $J_f$ and $J_b$ leading to $\eta_m$ for this case are shown in Fig.~\ref{fig_hBN_Cu}(e), while the stationary temperatures of the intermediate body in the two scenarios are represented in Fig.~\ref{fig_hBN_Cu}(f). Also in this case we observe a similar value for $T_f$ and $T_b$ in configurations yielding maximum rectification.

We stress that the temperature drop inside the intermediate body induced by the Kapitza resistance can be neglected. Indeed, at the interface between a polar material and a metal, the boundary thermal conductance $h_K$ takes typical values of $10^8$\,W\,m$^{-2}$\,K$^{-1}$~\cite{Kapitza}, so the corresponding Kapitza resistance $R_K$ is about $10^{-8}\,$m$^2$\,K\,W$^{-1}$. With the heat flux level we get in our device at the considered operating temperatures ($J\sim 10^4$\,W\,m$^{-2}$), the temperature drop $\Delta T$ at the interface between the materials $M_1$ and $M_2$ is $\Delta T= R_K J\sim10^{-4}\,$K. Moreover, the thermal conductivity in both materials ($\kappa_\mathrm{hBN}=30$\,W\,m$^{-1}$\,K$^{-1}$ and $\kappa_\mathrm{Cu}=400$\,W\,m$^{-1}$\,K$^{-1}$) leads to a temperature variation through the intermediate body of the order of $10^{-4}\,$K with layers $1\,\mu$m thick, so this variation can be neglected as well.

\section{Conclusions\label{sec:conclusions}}

We have introduced a rectification mechanism for the near-field radiative heat transport in many body-systems. Unlike the classical thermal rectification in two-terminal systems which require a noticeable temperature dependence of the optical properties of the materials, here the asymmetry in the heat transport is induced by many-body interactions. Moreover, we have shown that this mechanism can rectify heat fluxes even without a temperature dependence of the optical properties.
By investigating this effect in a three-body system, we have demonstrated that the existence of a privileged direction for the heat transport results from the interplay between the asymmetric coupling of the passive intermediate body with the external terminals and the temperature reached by this body in the stationary state.
Rectification coefficients larger than 50\% have been highlighted in near-field regime with simple systems made with polar dielectrics and metallic layers. A thermo-optical optimization of these systems should increase this coefficient.
These thermal rectifiers are fundamentally different from the previous ones. They do not require the use of phase-change materials and they can be designed to operate over a broad temperature range. Hence, these devices could find wide application in the field of thermal management and thermal regulation at the nanoscale.

\begin{acknowledgments}
This project has received funding from the European Union’s Horizon 2020 research and innovation programme under the Marie Sklodowska-Curie grant agreement No~892718~(I.L.).  
\end{acknowledgments}

\appendix*

\section{Optical reflection and transmission coefficients}

Here we give the single-body optical reflection and transmission coefficients which are required to compute the energy transmission coefficients.
For bodies 1 and 3, which are assumed semi-infinite and homogeneous, we have
\begin{align}
\rho_1=r_{01}^p,\qquad \rho_3=r_{02}^p,
\end{align}
where $r_{ab}^p$ is the Fresnel reflection coefficient of the interface between media $a$ and $b$, 
\begin{equation}
r^\mathrm{TE}_{ab}=\frac{k_{za}-k_{zb}}{k_{za}+k_{zb}},\qquad 
r^\mathrm{TM}_{ab}=\frac{\varepsilon_b k_{za}-\varepsilon_a k_{zb}}{\varepsilon_b k_{za}+\varepsilon_a k_{zb}},
\end{equation}
in which $a=0$ refers to vacuum and $a=i$ to material $M_i$. Here $\varepsilon_a$ is the complex permittivity of medium $a$ and $k_{za}=\sqrt{(\omega/c)^2 \varepsilon_a- k^2}$ is the normal component of the wave vector inside this medium. To describe the scattering properties of the intermediate body, we also introduce the Fresnel transmission coefficients $t_{ab}^p$ of the interface between media $a$ and $b$ which are given by
\begin{equation}
t^\mathrm{TE}_{ab}=\frac{2k_{za}}{k_{za}+k_{zb}},\qquad 
t^\mathrm{TM}_{ab}=\frac{2\sqrt{\varepsilon_a \varepsilon_b} k_{za}}{\varepsilon_b k_{za}+\varepsilon_a k_{zb}}.
\end{equation}
The intermediate body is a bilayer and has three interfaces, so that
\begin{align}
\rho_2^+&= r^p_{02} +\frac{t_{02}^p t_{20}^p  r^p_{210} e^{i2k_{z2}\delta_2}}{1+r^p_{02}r^p_{210}e^{i2k_{z2}\delta_2}},\\
\rho_2^-&= r^p_{01} +\frac{t_{01}^p t_{10}^p  r^p_{120} e^{i2k_{z1}\delta_1}}{1+r^p_{01}r^p_{120}e^{i2k_{z1}\delta_1}},
\end{align}
where $r^p_{abc}$ are the reflection coefficients of the two-interface system constituted by the interfaces between media $a$ and $b$ and between media $b$ and $c$, which here take the form
\begin{align}
r^p_{210} &= r^p_{21} + \frac{t_{21}t_{12}r^p_{10}e^{i2k_{z1}\delta_1}}{1+ r_{21} r_{10} e^{i2k_{z1}\delta_1}},\\
r^p_{120} &= r^p_{12} + \frac{t_{12}t_{21}r^p_{20}e^{i2k_{z2}\delta_2}}{1+ r_{12} r_{20} e^{i2k_{z2}\delta_2}}.
\end{align}
Lastly, the optical transmission coefficient of the intermediate body reads
\begin{equation}
\tau_2=\frac{t_{01}^pt_{120}^pe^{ik_{z1}\delta_1}}{1+r_{01}^pr_{120}^pe^{i2k_{z1}\delta_1}} ,
\end{equation}
where $t_{abc}^p$ is the two-interface transmission coefficient for which
\begin{equation}
t_{120}^p= \frac{t_{12}^p t_{20}^p e^{ik_{z2}\delta_2}}{1 +r_{12}^p r_{20}^p e^{i2k_{z2}\delta_2}}.
\end{equation}

\end{document}